\documentclass[twocolumn,english,aps,prd,nofootinbib]{revtex4-1}
\usepackage{lmodern}
\usepackage[T1]{fontenc}
\usepackage[latin9]{inputenc}
\usepackage{xcolor}
\usepackage{babel}
\usepackage{textcomp}
\usepackage{amsmath}
\usepackage{amsthm}
\usepackage{amssymb}
\usepackage{graphicx}
\PassOptionsToPackage{normalem}{ulem}
\usepackage{ulem}
\usepackage[unicode=true]
{hyperref}

\makeatletter

\theoremstyle{plain}

\theoremstyle{remark}

\allowdisplaybreaks
\newcommand{\bq}{\begin{equation}}
	\newcommand{\eq}{\end{equation}}
\newcommand{\bqn}{\begin{eqnarray}}
	\newcommand{\eqn}{\end{eqnarray}}
\newcommand{\nb}{\nonumber}
\newcommand{\lb}{\label}

\usepackage{enumerate}

\makeatother

\providecommand{\remarkname}{Remark}
\providecommand{\theoremname}{Theorem}

\begin{document}

\title{Observational test of ${\cal R}^{2}$ spacetimes with the S2 star in the Milky Way galactic center}

\author{Jian-Ming Yan$\,$}
\email{yanjm@zjut.edu.cn}
\affiliation{Institute for Theoretical Physics \& Cosmology, Zhejiang University of Technology, Hangzhou, 310023, China}
\affiliation{United Center for Gravitational Wave Physics, Zhejiang University of Technology, Hangzhou, 310023, China}

\author{Tao Zhu$\,$}
\email{zhut05@zjut.edu.cn; Corresponding author}
\affiliation{Institute for Theoretical Physics \& Cosmology, Zhejiang University of Technology, Hangzhou, 310023, China}
\affiliation{United Center for Gravitational Wave Physics, Zhejiang University of Technology, Hangzhou, 310023, China}

\author{Mustapha Azreg-A\"inou$\,$}
\email{azreg@baskent.edu.tr}
\affiliation{Ba\c{s}kent University, Engineering Faculty, Ba\u{g}lica Campus, 06790-Ankara, Turkey}

\author{Mubasher Jamil}
\email{mjamil@sns.nust.edu.pk}
\affiliation{School of Natural Sciences, National University of Sciences and Technology, H-12, Islamabad, 44000, Pakistan}

\author{Hoang Ky Nguyen$\,$}
\email{hoang.nguyen@ubbcluj.ro}
\affiliation{Department of Physics, Babe\c{s}--Bolyai University, Cluj-Napoca 400084, Romania}

\date{\today}

\begin{abstract}

A novel class of vacuum metrics expressible in analytical form was recently found for pure $\mathcal R^2$ gravity, based on a groundwork put forth by Buchdahl in 1962. These Buchdahl-inspired solutions offer a practical framework for testing ${\cal R}^2$ gravity through empirical observations. Within a subclass of asymptotically flat Buchdahl-inspired vacuum spacetimes, we identified a parameter $\epsilon$ measuring the deviation from the classic Schwarzschild metric, which corresponds to $\epsilon=0$. In this paper, we employ observational data from the S2 star's orbit around Sgr A* in the Milky Way galactic center and perform Monte Carlo Markov Chain simulations to probe the effects of the new metrics on the orbit of the S2 star. Our analysis presented herein reports a range at 95\% confidence level on the deviation parameter as $\epsilon\in(-0.6690,\ 0.4452)$. While no decisive evidence either in favor or in disfavor of the asymptotically flat Buchdahl-inspired spacetimes has been achieved, the obtained bound is compatible with the tighter results using other data of different nature as recently reported in Eur.\,Phys.\,J.\,C $\bf 84$, 330 (2024). As a meaningful test probing into a strong-field regime, our present study calls for further observations with prolonged period and improved accuracy in order to tighten the bound for $\epsilon$ using the S2 star orbit.

\end{abstract}

\maketitle

\section{Introduction}

In the past hundred years, Einstein's General Relativity (GR) has proven to be a very successful gravitational theory, precisely delineating the dynamics of massive objects subjected to gravity. It provides accurate descriptions for a wide range of empirical phenomena, such as the precession of Mercury's perihelia, the gravitational redshifts, and the Shapiro time delay. Even more remarkably, it produces powerful predictions of novel gravitational entities like black holes, compact stars, and gravitational waves. In the more recent decades, meticulous observations and analyses of black hole shadows at the centers of M87 and the Milky Way galaxies have been conducted \cite{EHT1, EHT2, EHT3, EHT4}. Simultaneously, numerous gravitational wave signals, originating from the mergers of binary black holes with differing masses, have been detected \cite{gws, gws2}. GR has consistently demonstrated its accuracy in these tests, solidifying its status as the only gravity theory to have successfully passed all solar system and astronomical examinations.

However, despite its remarkable successes, GR grapples with some fundamental challenges. These include the need for renormalization and the imperative task of establishing consistency with quantum mechanics. This quest aims to formulate a comprehensive set of physical laws applicable across various length scales, encompassing the very small and the exceptionally large scales.

These shortcomings indicate the need for modifications beyond GR. In exploring various theoretical frameworks of quantum gravity, string theory has been proposed as a unified theory of relativity and quantum mechanics. Its effective string action not only includes the Einstein-Hilbert action but also encompasses a series of high-order curvature correction terms. Research on the specific gravity model ${\cal R}+{\cal R}^{2}$ has demonstrated that ${\cal R}^{2}$ gravity exhibits scale invariance, lacks ghost problems, is renormalizable, and can be embedded into supergravity \cite{Gaume-2016}. Thus, this simplified model is considered a reasonable candidate for modified gravity and quantum gravity. However, the task of analytically solving the gravitational equations for ${\cal R}^{2}$ is quite challenging and largely remains unsuccessful  due to the action of ${\cal R}^{2}$ gravity leads to field equations involving fourth-order derivatives of the metric tensor with respect to the metric tensor variations.

In previous research, Buchdahl endeavored to discover new static and spherically symmetric spacetime or black hole solutions within the framework of the ${\cal R}^{2}$  theory \cite{Buchdahl-1962}. His work ultimately formulated a second-order ordinary differential equation aimed at solving long-standing unresolved metric coefficients. Until recently, Nguyen achieved success in obtaining vacuum solutions \cite{Nguyen-2022-Buchdahl, Nguyen-2022-Lambda0, Nguyen-2023-essay, 2023-WH, 2023-axisym, 2023-WEC}. For solutions that are asymptotically flat, the resulting metric involves a free parameter, the Buchdahl parameter $k$, a higher-derivative characteristic. When setting $k = 0$, one can recover the Schwarzschild spacetime/black hole.

Naturally, there is a desire to subject this new solution to some phenomenological tests. Currently, the prevailing consensus suggests that Sgr A* at the center of the Milky Way is a supermassive black hole, with the S2 star exhibiting periodic orbital motion around this black hole. Utilizing observational data from S2 star, extensive tests on gravitational effects have been conducted \cite{deMartino:2021daj, DellaMonica:2021xcf, DellaMonica:2021fdr, DAddio:2021smm, nohair, nohair2, Benisty:2021cmq, Han:2014yga, Yan:2023vdg, Yan:2022fkr, Zhang:2024fpm, DeMartino:2023qkl, DellaMonica:2023dcw, deLaurentis:2022oqa, Fernandez:2023kro, Cadoni:2022vsn, Shaymatov:2023jfa}, and the GRAVITY collaboration has measured the orbital precession of S2 using various datasets \cite{GRAVITY2}. We also aim to leverage publicly available observational data of the S2 star \cite{Mon} to test the gravitational effects under the asymptotically flat Buchdahl-inspired metric and compare it with theoretical predictions. We would like to point out that recently we calculated the effects of Buchdahl-inspired spacetimes on astronomical observations considering both within and outside of the solar system, including the deflection angle of light by the Sun, gravitational time delay, perihelion advance of S2 star, shadow, and geodetic precession \cite{taohoang}. Furthermore, we determined observational constraints on the corresponding deviation parameters by comparing theoretical predictions with the most recent observations. Among these constraints, we find that the tightest one comes from the Cassini mission's measurement of gravitational time delay \cite{taohoang}.

In this paper, we analyze the experimental and observational implications of the asymptotically flat Buchdahl-inspired metric. We aim to investigate the extent to which new effects, beyond the Schwarzschild scenario, influence the dynamics of particles in geodesic motion. More generally, we consider an extension of the asymptotically flat Buchdahl-inspired metric, described by several independent parameters. By calculating the dynamics of particles in a two-body system, we perform a Monte Carlo Markov Chain (MCMC) analysis to test our predictions with the observational data of S2 star orbiting around Sgr A* at the center of the Milky Way.

The remainder of this paper is organized as follows: Section II offers a succinct introduction to the Buchdahl-inspired metric, followed by an examination of the geodesic equations governing the motion of massive objects within this spacetime. Leveraging these equations, we meticulously derive the impacts on the dynamics of particles undergoing geodesic motion. In Section III, we present the dataset utilized in our MCMC analysis, encompassing positional and velocity data for the S2 star. Section IV then presents the outcomes derived from the MCMC analysis and compares them with theoretical predictions. Finally, Section V succinctly summarizes our findings and initiates a discussion. Moreover, Section VI provides additional insights and perspectives on our results. 

Throughout this paper, we adopt units where $c = 1$.

\section{GEODESICS AND  MOTION OF PARTICLES IN THE GENERAL BUCHDAHL-INSPIRED METRICS}

In this section, we offer a concise introduction to geodesics within the broader class of Buchdahl-inspired metrics. Subsequently, we derive the equations of motion governing the orbital dynamics of massive objects within this metric framework. By examining the evolution of geodesics, we can calculate various observational quantities. This enables us to employ observational data to constrain and refine our understanding of these metrics.

\subsection{The general Buchdahl-inspired metrics}

In this subsection, we present the general asymptotically flat Buchdahl-inspired metrics in standard coordinates and examine the geodesics of massive objects within this metric. Using coordinates $(t, R, \theta, \phi)$, the metric can be cast in the Morris-Thorne form as follows \cite{MorrisThorne}:
\begin{eqnarray}
	ds^{2} & = & -e^{2\Phi(R)}dt^{2}+\frac{dR^{2}}{1-\frac{b(R)}{R}}+R^{2}d\Omega^{2}.\label{R_coors}
\end{eqnarray}
with funtions $\Phi(R)$ and $b(R)$ given in \cite{2023-WH}.  Briefly speaking, the areal radius $R$ is expressed in terms of an auxiliary coordinate $y$ per
\begin{equation}
R=\zeta\,r_{\text{s}}\frac{y^{\frac{\tilde{k}-1}{\zeta}+1}}{1-y^2}\label{eq:areal-radius}
\end{equation}
The redshift function and the shape function, respectively, are
\begin{align}
e^{2\Phi(R)} & = y^{\frac{2}{\zeta}(\eta+1)},\label{eq:redshift-func}\\
1-\frac{b(R)}{R} & = \frac{1}{4y}\left[(y^2+1)+\frac{\tilde{k}-1}{\zeta}(1-y^2)\right]^{2}\label{eq:shape-func}
\end{align}
For $\eta=\tilde k,\ \zeta=\sqrt{1+3\tilde{k}^2}$, the metric~\eqref{R_coors}--\eqref{eq:shape-func} produces the asymptotically flat Buchdahl-inspired metric, derived in~\cite{Nguyen-2022-Lambda0}. For $\zeta=1$, it recovers the Campanelli--Lousto metric proposed in~\cite{CampanelliLousto-1993}.

At the typical radius of the S0-2 star's orbit, the gravitational potential due to the gravity of the supermassive black hole is roughly $~10^{-5}$, thus the weak field limit is valid. Under the weak field approximation, this metric can be approximated as:
\begin{eqnarray}
ds^{2}&\simeq&-\left(1-\frac{(1+\eta)r_{s}}{R}+\frac{(1+\eta)(\tilde{k}+\eta)r_{s}^{2}}{2R^{2}}\right)dt^{2}\nonumber \\
&& +\left(1+\frac{(1-\tilde{k})r_{s}}{R}\right)dR^{2}+R^{2}d\Omega^{2}.\label{weak}
\end{eqnarray}
Here, $\eta$ and $\tilde{k}$ are treated as two independent parameters arising from the Buchdahl-inspired metric, and the radius $r_s$ is related to the mass $M$ of the solution through $r_s=2 G M$, and $G$ is the gravitational constant. 
Comparing this weak field expansion with the Newtonian limit, one can establish a connection between $G$ and the Newtonian gravitational constant $G_{\rm N}$ as $(1+\eta)G=G_{\rm N}$. Therefore, one can get:
\begin{eqnarray}
g_{tt}=1-\frac{2 G_{\rm N} M}{R}+2\frac{\tilde{k}+\eta}{1+\eta}\frac{G_{ \rm N }^{2}M^{2}}{R^{2}}\equiv f(R).
\end{eqnarray}
To simplify the calculation, we define a deviation parameter $\epsilon$, which characterizes the deviation from the Schwarzschild black hole, as follows:
\begin{eqnarray}
\epsilon \equiv \frac{\tilde{k}+\eta}{1+\eta}, \quad G_{\text{N}}=1.
\end{eqnarray}
Then the metric \eqref{weak} becomes:
\bqn
ds^{2}&\simeq&-\left(1-\frac{2 M}{R}+\epsilon \frac{2 M^{2}}{R^{2}}\right)dt^{2}\nonumber \\
&& +\left(1+\frac{2(1-\epsilon)M}{R}\right)dR^{2}+R^{2}d\Omega^{2}.\label{newton}
\eqn

\subsection{Equations of motion of massive objects in the general Buchdahl-inspired metric}

To study the motion of massive objects in the Buchdahl-inspired metric  (\ref{newton}), we can view it as a perturbation of a Keplerian elliptical orbit.  One can  transform the coordinates $(R, \theta, \phi)$ into the isotropic coordinates $(r, \theta, \phi)$.  With the coordinates $(r, \theta, \phi)$, the metric of the Buchdahl-inspired metric is written as
\begin{eqnarray}
ds^{2}\simeq -f(r)dt^{2}+g(r)\Big[dr^{2}+r^{2}(d\theta^{2}+\sin^{2}\theta d\phi^{2})\Big],\label{iso}
\end{eqnarray}
where
\begin{eqnarray}
	f(r)&=& 1-\frac{2M}{r}+\frac{2M^2}{r^2}, \\
	g(r)&=& 1+\frac{2(1-\epsilon)M}{r}.
\end{eqnarray}
When considering the motion of a massive object following a time-like geodesic in the Buchdahl-inspired metric, we have
\begin{eqnarray}
\frac{d^{2}x^{\mu}}{d\tau^2}+\Gamma ^{\mu}_{~\nu \rho} \frac{dx^{\nu}}{d\tau}\frac{dx^{\rho}}{d\tau}=0,\label{geodesic}
\end{eqnarray}
where \(\tau\) is the proper time of the massive object, and \(\Gamma ^{\mu}_{~\nu \rho}\) represents the Christoffel symbols of the Buchdahl-inspired metric.

For the specific case of the S2 star orbiting around Sgr A* in the galactic center, we find it convenient to consider the weak field approximation. In this context, we can derive the equations of motion for massive particles by expanding the geodesic equation \eqref{geodesic} in terms of small quantities, namely \(M\) and \(\boldsymbol{v}\), where \(\boldsymbol{v}\) represents the velocity of the massive particles.

By performing this expansion, the geodesic equation can be transformed into a perturbed Kepler problem in celestial mechanics, described by:
\begin{eqnarray}
\frac{d^2\boldsymbol{r}}{dt^2} = -\frac{M}{r^3}\boldsymbol{r} + \boldsymbol{F}_{\text{pert}},\label{pert}
\end{eqnarray}
where \(\boldsymbol{r}\) is the position vector, \(t\) is time, \(M\) is the mass of the central object (Sgr A*), \(r\) is the distance between the massive object and the central object, and \(\boldsymbol{F}_{\text{pert}}\) represents the perturbing forces arising from the Buchdahl-inspired metric.

With the Christoffel symbols calculated for the Buchdahl-inspired metric, as used in \eqref{geodesic}, one obtains
\begin{eqnarray}
	\frac{d^2 \boldsymbol{r} }{dt^2} &=& -\frac{M}{r^2}\frac{\boldsymbol{r}}{r}+\frac{(4-2\epsilon)M^2}{r^3}\frac{\boldsymbol{r}}{r} \nonumber\\
	&&+\frac{(\epsilon -1) M \boldsymbol{v^2}}{r^2} \frac{\boldsymbol{r}}{r}+\frac{(4-2 \epsilon)M}{r^3}(\boldsymbol{r} \cdot \boldsymbol{v}) \boldsymbol{v}.
\end{eqnarray}
By comparing with \eqref{pert}, it is easy to find 
\begin{eqnarray}
\begin{aligned}
	\boldsymbol{F}_{\text{pert}} =& \frac{(4-2\epsilon)M^2}{r^3}\frac{\boldsymbol{r}}{r}\\
	&+\frac{(\epsilon -1) M \boldsymbol{v^2}}{r^2}\frac{\boldsymbol{r}}{r} +\frac{(4-2 \epsilon)M}{r^3}(\boldsymbol{r} \cdot \boldsymbol{v}) \boldsymbol{v}.
\end{aligned}
\end{eqnarray}

\begin{figure}
\centering
\includegraphics[width=8.1cm]{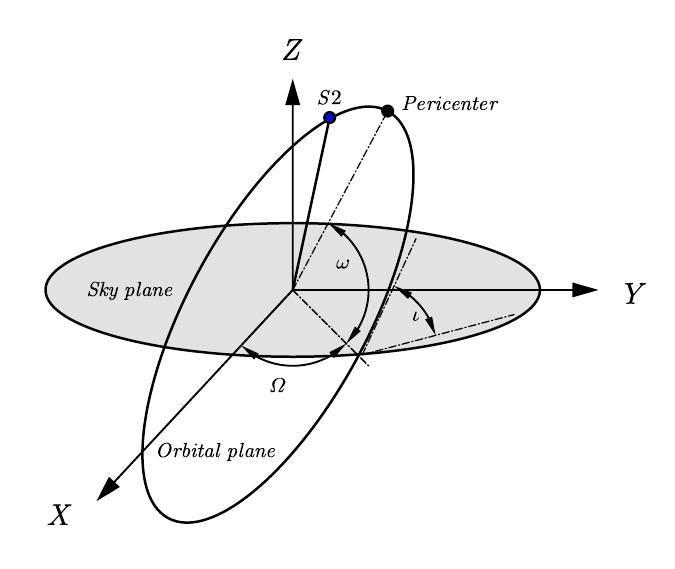}
\caption{This figure illustrates the coordinate system used to describe the orbits of stellar stars in the galactic center. The $Z$-axis aligns with the line of sight, and the coordinates $[X, Y] = [\text{Ra(S2)} - \text{Ra(Sgr A*)}, \text{Dec(S2)} - \text{Dec(Sgr A*)} ]$. Here, $\omega$ represents the pericenter argument, $\Omega$ denotes the longitude of the ascending node, and $\iota$ represents the orbital inclination of the precessing elliptical orbit for the S2 star. } \label{cor}
\end{figure}

To study the effect of the perturbed force $\boldsymbol{F}_{\text{pert}}$, we project $\boldsymbol{F}$ into three directions $(\boldsymbol{ e_{r}}, \boldsymbol{ e_{\theta}}, \boldsymbol{ e_{z}})$, 
\begin{eqnarray}
\boldsymbol{F}=\mathcal{D}\boldsymbol{ e_{r}}+\mathcal{S}\boldsymbol{e_{\theta}}+\mathcal{W}\boldsymbol{e_{z}},
\end{eqnarray}
where $\boldsymbol{ e_{r}}$  along the direction of $\dot{r}$, $\boldsymbol{e_{\theta}}$  along the direction of $\dot{\theta}$,  $\boldsymbol{e_z}$ along the direction of  $\boldsymbol{e_r \times e_{\theta}}$, and $ \boldsymbol{r}=r\boldsymbol{e_r} $, $\boldsymbol{v}= \dot{r}\boldsymbol{e_r}+r\dot{\theta}e_{\theta}$.  The three components ${\cal D}$, ${\cal S}$, and ${\cal W}$ are given by 
\begin{eqnarray}
\begin{aligned}
	\mathcal{D}=&\frac{(4-2 \epsilon)M^2}{r^3}+ \frac{(\epsilon-1)M \boldsymbol{v^2}}{r^2} +\frac{(4-2\epsilon)M}{r^{3}}(\boldsymbol{r}\cdot\boldsymbol{v})\dot{r},\\
	\mathcal{S}=&\frac{(4-2\epsilon)M}{r^{3}}(\boldsymbol{r}\cdot\boldsymbol{v})r\dot{\theta},\\
	\mathcal{W}=&0. \label{RSW}
\end{aligned}
\end{eqnarray}
For such a  Keplerian elliptical orbit in Fig. \ref{cor}, we have
\begin{eqnarray}
\begin{aligned}
	r=&\frac{a(1-e^2)}{1+e\cos (\theta-\omega)},\\
	\dot{r}=&\sqrt{\frac{M}{p}}e\sin (\theta-\omega),\\ 
	\dot{\theta}=&\sqrt{\frac{M}{p^3}}\Big[1+e\cos(\theta-\omega)\Big].
\end{aligned}
\end{eqnarray}
In the given context, where \(a\) represents the semi-major axis, \(e\) is the eccentricity, \(\omega\) is the argument of pericenter, \(\dot{r}\) denotes the radial velocity, \(\dot{\theta}\) represents the angular velocity, and \(p=a(1-e^2)\) stands for the semi-latus rectum of the elliptic orbit, we can further elaborate on these parameters:

\begin{itemize}
\item \(a\): The semi-major axis is a fundamental parameter that defines the size of the ellipse. It represents half of the major axis, which is the longest diameter of the elliptic orbit.

\item \(e\): Eccentricity measures how much the shape of the orbit deviates from a perfect circle. For \(0 \leq e < 1\), the orbit is elliptical; for \(e = 0\), it is a perfect circle; and for \(e > 1\), the orbit is hyperbolic.

\item \(\omega\): The argument of pericenter describes the orientation of the ellipse in its orbital plane. It specifies the angle between the ascending node and the pericenter, providing information about the location of the closest approach to the central body.

\item \(\dot{r}\): Radial velocity is the rate of change of the distance between the object and the central body along the radial direction. It determines how fast the object is moving towards or away from the central body.

\item \(\dot{\theta}\): Angular velocity is the rate at which the angular position of the object changes with respect to time. It characterizes the rotational motion of the object around the central body.

\item \(p\): The semi-latus rectum is a parameter related to the shape of the orbit. It is defined as \(p = a(1-e^2)\) and is crucial in determining various properties of the elliptic orbit.
\end{itemize}

Considering the effect perturbation force $\boldsymbol{F_{\rm pert}}$, we can use the following dynamical equations to study the evolution of  the five orbital parameters $a, e, \omega, \Omega, \iota$, given by \cite{willbook}
\bqn
\frac{da}{dt}&=&2\sqrt{\frac{a^{3}}{M}}(1-e^{2})^{-\frac{1}{2}}\Big[e\sin f \mathcal{D}+(1+e\cos f)\mathcal{S}\Big],\\ \lb{da}
\frac{de}{dt}&=&\sqrt{\frac{a}{M(1-e^{2})}}\Big[\sin f \mathcal{D} \nb\\ 
&&~~~~ + \frac{2\cos f +e(1+\cos^{2}f) }{1+e\cos f}\mathcal{S}\Big],\\ \lb{de}
\frac{d\iota}{dt}&=&\sqrt{\frac{a}{M(1-e^{2})}}\frac{\cos (\omega + f)}{1+e \cos f}\mathcal{W},\\ \lb{di}
\frac{d\Omega}{dt}&=&\sqrt{\frac{a}{M(1-e^{2})}}\frac{\sin (\omega +f)}{\sin \iota (1+e\cos f) }\mathcal{W},\\ \lb{dO}
\frac{d\omega}{dt}&=&\frac{1}{e}\sqrt{\frac{a}{M(1-e^{2})}}\Big[-\cos f \mathcal{D}+\frac{2+e\cos f}{1+e \cos f}\sin f \mathcal{S} \nb\\  
&&~~~~ -e\cot \iota \frac{\sin (\omega +f)}{1+e\cos f}\mathcal{W}\Big],\\ \lb{do}
\frac{df}{dt}&=&\sqrt{\frac{M(1-e^{2})}{a^{3}}}\Big(1+e\cos f\Big)^2 \nb\\  
&& +\frac{1}{e}\sqrt{\frac{a}{M(1-e^{2})}}\Big[\cos f \mathcal{D}-\frac{2+e\cos f}{1+e\cos f}\sin f\mathcal{S}\Big], \nb\\ 
\eqn
where $\iota$ is the orbital inclination, $f=(\theta-\omega)$ is the true anomaly of the elliptic orbit. Upon integrating these equations, one can obtain the orbit of such a massive object in the Buchdahl-inspired metric. Furthermore, these equations can be transformed to \cite{willbook}
\begin{eqnarray}
\begin{aligned}
	\frac{da}{df} \approx&  2\frac{p^{3}}{M}\frac{1}{(1+e\cos f)^{3}}\mathcal{S},\\
	\frac{de}{df} \approx& \frac{p^{2}}{M}\Big[\frac{\sin f}{(1+e \cos f)^2}\mathcal{D} \\
	& +\frac{2\cos f +e(1+\cos^{2}f)}{(1+e\cos f)^{3}}\mathcal{S}\Big],\\
	\frac{d \iota}{df}\approx& \frac{p^2}{M}\frac{\cos (\omega+f)}{(1+e\cos f)^3}\mathcal{W},\\
	\frac{d\Omega}{df}\approx&\frac{p^2}{M}\frac{\sin (\omega+f)}{\sin \iota(1+e\cos f)^3}\mathcal{W},\\
	\frac{d\omega}{df}\approx&\frac{1}{e}\frac{p^2}{M}\Big[-\frac{\cos f}{(1+e\cos f)^2}\mathcal{D}+\frac{2+e\cos f}{(1+e\cos f)^3}\sin f \mathcal{S} \\
	& -e\cot \iota \frac{\sin (\omega+f)}{(1+e\cos f)^3}\mathcal{W}\Big].
\end{aligned}
\end{eqnarray}
The secular changes of the five orbital elements $a, e, \omega, \Omega, \iota$ can be calculated via
\begin{eqnarray}
\dot{\mu}^{\alpha}=\frac{1}{T}\int_{0}^{T}\frac{d\mu^{\alpha}}{dt}dt\simeq \frac{1}{T}\int_{0}^{2\pi}\frac{d\mu^{\alpha}}{df}df,
\end{eqnarray}
where $T=2\pi\sqrt{a^3/M}$ is the period of an elliptical orbit. 

Then performing the above integrals for each orbital element, one obtains the drift rates of the five orbital elements,
\bqn
\dot a &=&0, \\
\dot e &=&0, \\
\dot \iota &=&0, \\
\dot \Omega &=&0, \\
\dot \omega &=& \frac{3 M^{3/2}}{a^{5/2} (1-e^2)} \left(1- \frac{2}{3}\epsilon\right).
\eqn
These results show explicitly that only the argument of pericenter $\omega$ of the elliptical orbit changes over time. This effect causes the precession of the pericenter of the massive objects orbiting around the Buchdahl-inspired spacetime. The precession rate per orbit reads
\begin{eqnarray}
\Delta \omega =\frac{6\pi M}{a(1-e^2)}\Big(1-\frac{2}{3} \epsilon \Big).\label{precession}
\end{eqnarray}
The expression for $\Delta \omega$ indicates a linear dependence on the parameter $\epsilon$. As $\epsilon$ varies, the correction term in~\eqref{precession} to the precession rate experiences a proportional change. Notably, when $\epsilon = 0$, the precession rate reverts to the well-known Schwarzschild result. This linear relationship underscores the influence of the Buchdahl parameter on the orbital precession, offering insights into how deviations from the Schwarzschild case impact the dynamics of celestial bodies in this gravitational framework.

Based on the measurement of S2's orbital precession by the GRAVITY Collaboration \cite{GRAVITY2}, given by
\bqn
\Delta\phi_{\rm per\;orbit}=12.1\times(1.10 \pm 0.19) ,\lb{pre}
\eqn
one can estimate the value of $\epsilon$ to be
\bqn
-0.42 < \epsilon < 0.134. \label{prediction}
\eqn

\section{DATA AND DATA ANALYSIS OF THE S0-2 STAR IN THE GALACTIC CENTER}
\begin{figure*}
\centering
\includegraphics[width=13.1cm]{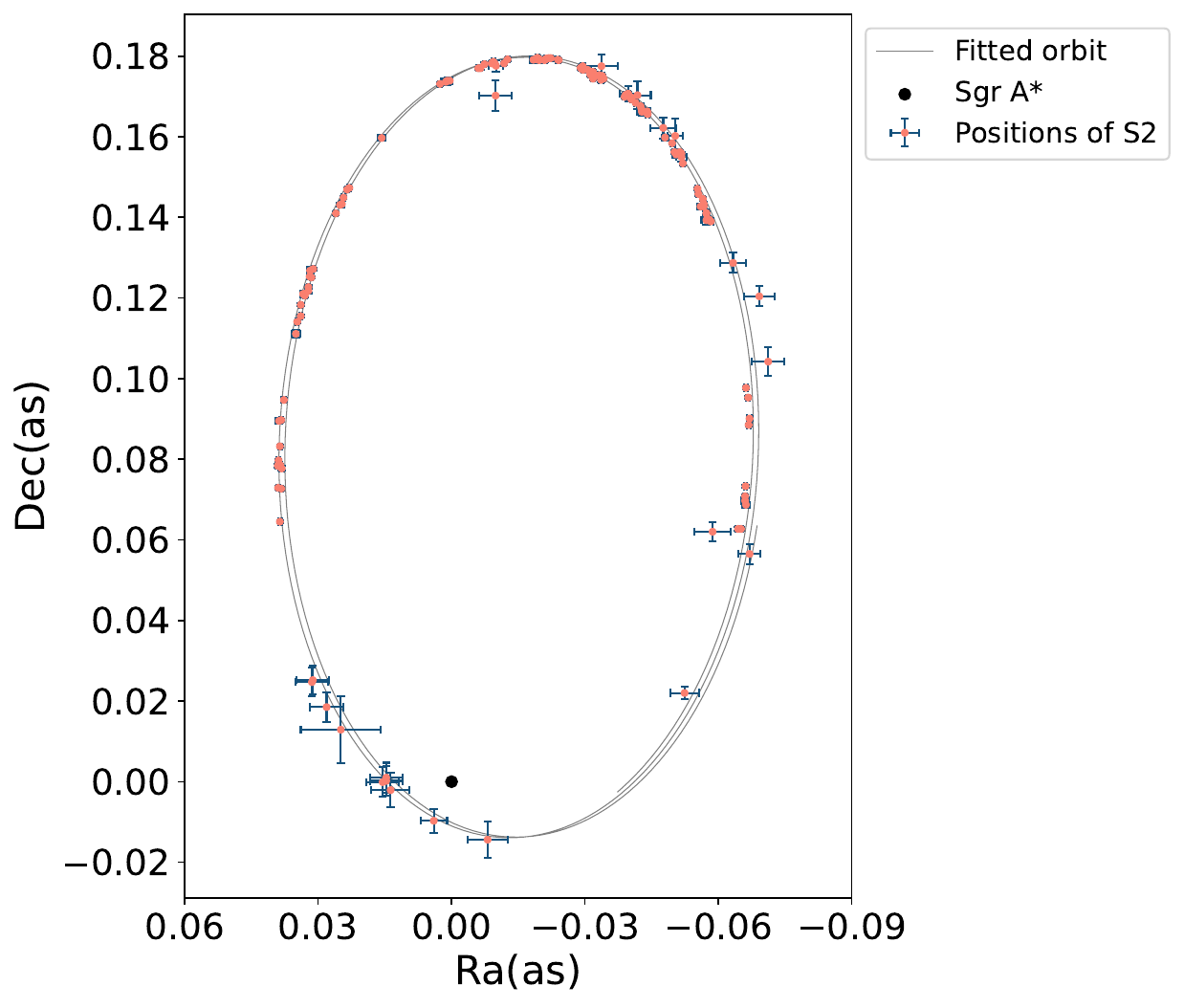}
\caption{ 
The pink points represent the astrometric positions of S2 on the sky plane, with the blue bars indicating the associated error margins. The grey line shows the fitted orbit of S2 as it revolves around the galactic center. The orbital precession is distinctly visible, highlighting S2's close proximity to the galactic center. }\label{orbit} 
\end{figure*}

\begin{figure*}
\centering
\includegraphics[width=13.1cm]{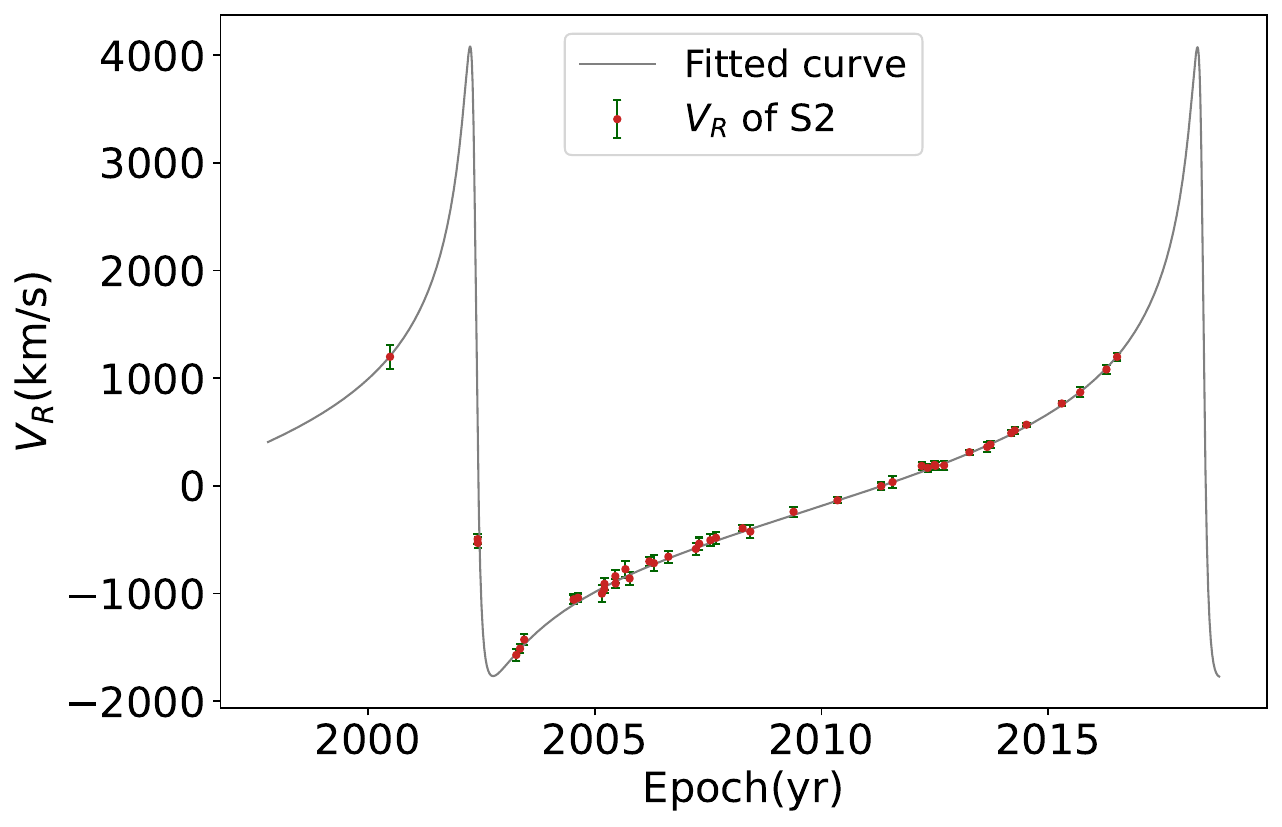}
\caption{ 
The red points represent the radial velocity of S2 in each epoch, with the green bars indicating the associated error margins. From the fitted curve, one can notice that there is a periodic change of the velocity.
}\label{velocity}
\end{figure*}

In this section, we present the publicly available data \cite{Mon} pertaining to the orbit of the S2 star around the supermassive black hole Sgr A* in the galactic center. The dataset consists of two essential components: astrometric positions, representing the star's projected locations on the sky plane, shown in Fig.~\ref{orbit}, and radial velocity, shown in Fig.~\ref{velocity}, indicating the velocity component along the line of sight.

\subsection{Astrometric positions on the sky plane}
The astrometric positions on the sky plane are represented by $[X_{\rm{astro}}(t_{\rm{obs}}), Y_{\rm{astro}}(t_{\rm{obs}})]$, where $t_{\rm{obs}}$ denotes the time of observation of the light emitted from S2. However, the theoretical positions $[x_{\rm{orbit}}(t_{\rm{em}}), y_{\rm{orbit}}(t_{\rm{em}})]$ calculated are on the orbital plane, with $t_{\rm{em}}$ representing the time when the light is emitted from S2. To compare astrometric and theoretical positions, we need to transform $t_{\rm{em}}$ to $t_{\rm{obs}}$ and project the theoretical positions onto the sky plane.

First, we mainly consider the Romer time delay witch is  due to the finite speed of light, which can be expressed as:
\bqn
t_{\rm{obs}}-t_{\rm{em}}=\frac{Z(t_{\rm{em}})}{c},
\eqn
where $c$ is the speed of light and  $Z(t_{\rm{em}})$  is the distance between the celestial object on the orbital plane and the sky plane.

Next, we project the theoretical positions onto the sky plane using the Thiele-Innes constants, which are given by the relationships:
\bqn
X&=&xB+yG, \lb{Xobs}\\
Y&=&xA+yF, \lb{Yobs}\\
Z&=&xC+yH. \lb{Zobs}
\eqn
Here, the Thiele-Innes constants are defined as:
\bqn
A&=&\cos \Omega \cos \omega -\sin \Omega \sin \omega \cos \iota, \lb{A}\\
B&=&\sin \Omega \cos \omega +\cos \Omega \sin\omega \cos \iota, \lb{B}\\
C&=&\sin \omega \sin \iota, \lb{C}\\
F&=&-\cos\Omega \sin \omega -\sin \Omega \cos \omega \cos \iota, \lb{F}\\
G&=&-\sin \Omega \sin \omega +\cos \Omega \cos \omega \cos \iota, \lb{G}\\
H&=&\cos \omega \sin \iota. \lb{H}
\eqn
These equations establish the transformation between the theoretical positions in the orbital plane and the observed positions on the sky plane. The Thiele-Innes constants are determined by the orbital elements and inclination of the S2 star's orbit.

\subsection{Radial velocity}

The velocity dataset comprises the radial velocity and its corresponding observational time. Accounting for the photon's frequency shift, denoted as $\zeta$, is crucial in analyzing the radial velocity. This shift is defined by
\[
\zeta = \frac{\Delta \nu}{\nu} = \frac{\nu_{\rm em} - \nu_{\rm obs}}{\nu_{\rm obs}} = \frac{V_{\rm R}}{c},
\]
where $\nu_{\rm em}$ is the frequency of the photon at the time of emission, $\nu_{\rm obs}$ is the observed frequency, and $V_{\rm R}$ is the radial velocity of the S-stars.

In our examination of the photon's frequency shift, we consider two relativistic effects: the Doppler shift $\zeta_{\rm D}$ and the gravitational shift $\zeta_{\rm G}$. They are defined as follows:
\[
\begin{aligned}
\zeta_{\rm D} &= \frac{\sqrt{1 - \frac{v_{\rm em}^2}{c^2}}}{1 - \mathbf{n} \cdot \mathbf{v}_{\rm em}}, \\
\zeta_{\rm G} &= \frac{1}{\sqrt{-g_{00}(t_{\rm em},x_{\rm orbit}(t_{\rm em}))}},
\end{aligned}
\]
where $v_{\rm em}$ is the velocity at the time of emission, and $\mathbf{n} \cdot \mathbf{v}_{\rm em}$ is the Newtonian line-of-sight velocity. The total frequency shift $\zeta$ is then given by:

\[
\zeta = \zeta_{\rm D} \cdot \zeta_{\rm G} - 1.
\]

\subsection{Orbital precession of S2}

The orbital precession of the S2 star was recently measured by the GRAVITY Collaboration \cite{GRAVITY2}, as shown in Eq.~(\ref{pre}). In contrast to measurements of S2's orbit using only the positions and velocities of the S2 star, this new measurement incorporated a substantial amount of fresh data, as detailed in \cite{GRAVITY2}. Some of this data may overlap with the S2 star's position and velocity data discussed in the above subsections. However, as clarified in ref.~\cite{GRAVITY2}, the accuracy of the orbital precession measurement primarily hinges on how well the zero-point offsets and drifts of the reference frame's center are constrained. This critical aspect relies on 75 measurements of Sgr A* flares from NACO AO data spanning 2003 to 2019. The inclusion of this new flare data allows for independent constraint of the NACO reference frame's zero-point. By combining this with the position and velocity data from ref.~\cite{Do:2019txf}, the GRAVITY Collaboration demonstrated consistency in the precession measurement \cite{GRAVITY2}. Therefore, in this paper, we can treat the direct detection of the S2's orbital precession as an independent measurement and directly use it independently in our likelihood analysis later. The potential double counting of overlapping dataset segments is expected to have a negligible impact on our analysis.

\section{ANALYSIS OF MONTE CARLO MARKOV CHAIN}

In this section, we employ the Python package \textit{emcee} \cite{emcee} to conduct a MCMC analysis, aiming to constrain the parameter $\epsilon$ arising from the Buchdahl-inspired metric. The parameters explored in the MCMC analysis include
\[
\{M, R, a, e, i, \omega, \Omega, T_{\rm P}, \alpha, \delta, v_{\rm \alpha}, v_{\rm \delta}, v_{\rm LSR}, \epsilon \},
\]
where $M$ denotes the mass of the central black hole in Sgr A*, and $R$ represents the distance between the Earth and the black hole. $T_{\rm P}$ refers to the time of the pericenter passage, which we choose as the initial point for the calculation. The five orbital elements $\{ a, e, \iota, \omega, \Omega \}$ describe the osculating elliptical orbits of the S2 star. The parameters $\{\alpha, \delta, v_{\rm \alpha}, v_{\rm \delta}\}$ represent the zero-point offsets and drifts of the reference frame's center and $v_{\rm LSR}$ is the correction of the radial velocity measurements to the local standard of rest (LSR). 

For the likelihood function, we first consider the position part $\log {\cal L}_{\rm P}$, which is defined as
\begin{equation}
\begin{aligned}
	\log {\cal L}_{\rm P} = &-& \frac{1}{2} \sum_{i} \frac{(X_{\rm data}^i -X_{\rm theory}^i)^2}{(\sigma^i_{X,{\rm data}})^2} \\
	&-&\frac{1}{2} \sum_{i} \frac{(Y_{\rm data}^i -Y_{\rm theory}^i)^2}{(\sigma^i_{Y,{\rm data}})^2},
\end{aligned}
\end{equation}
where $X_{\rm data}^i$ and $Y_{\rm data}^i$ are the measured positions of the star at epoch $i$, $X_{\rm theory}^i$ and $Y_{\rm theory}^i$ are the corresponding positions predicted by the Buchdahl-inspired metric, and $\sigma^i_{X,{\rm data}}$ and $\sigma^i_{Y,{\rm data}}$ are the uncertainties of these measured data. 

Next we consider the likelihood of the radial velocity part $\log {\cal L}_{\rm VR}$, which is defined as
\begin{equation}
\log {\cal L}_{\rm VR} =- \frac{1}{2} \sum_{i} \frac{(V_{\rm R, data}^i - V_{\rm R, theory}^i)^2}{(\sigma^i_{V_{\rm R, data}})^2},
\end{equation}
where $V_{\rm R, data}^i$ is the measured radial velocity of the star at time $i$, $V_{\rm R, theory}^i$ is the corresponding radial velocity predicted by the Buchdahl-inspired metric, and $\sigma^i_{V_{\rm R, data}}$ is the uncertainty in the measurement. 

Furthermore, to strengthen the constraints of the MCMC analysis, we decided to add the precession (\ref{pre}) of S2 measured by  the GRAVITY Collaboration \cite{GRAVITY2} into the likelihood.
Therefore we  define this part as
\begin{equation}
\log {\cal L}_{\rm Pre} = - \frac{1}{2} \frac{(\Delta \phi_{\rm per \ orbit}-\Delta \phi_{\rm theory})^2}{\sigma^2_{\Delta \phi, {\rm per \ orbit}}},
\end{equation}
where $\Delta \phi_{\rm per \ orbit}$ is the measured orbital precession of S2, $\Delta \phi_{\rm theory}$ is the corresponding orbital precession predicted by our model with the given value of $\epsilon$, and $\sigma^2_{\Delta \phi, {\rm per \ orbit}}$ is the uncertainty in the measurement.

By combining these likelihood functions, we can obtain the overall likelihood function $\log {\cal L}$ as:
\bqn
\log {\cal L} = \log {\cal L}_{\rm P} + \log {\cal L}_{\rm VR} + \log {\cal L}_{\rm Pre}.\lb{likelihood}
\eqn

\section{RESULTS}

\begin{figure*}[h]
\centering
\includegraphics[width=\textwidth]{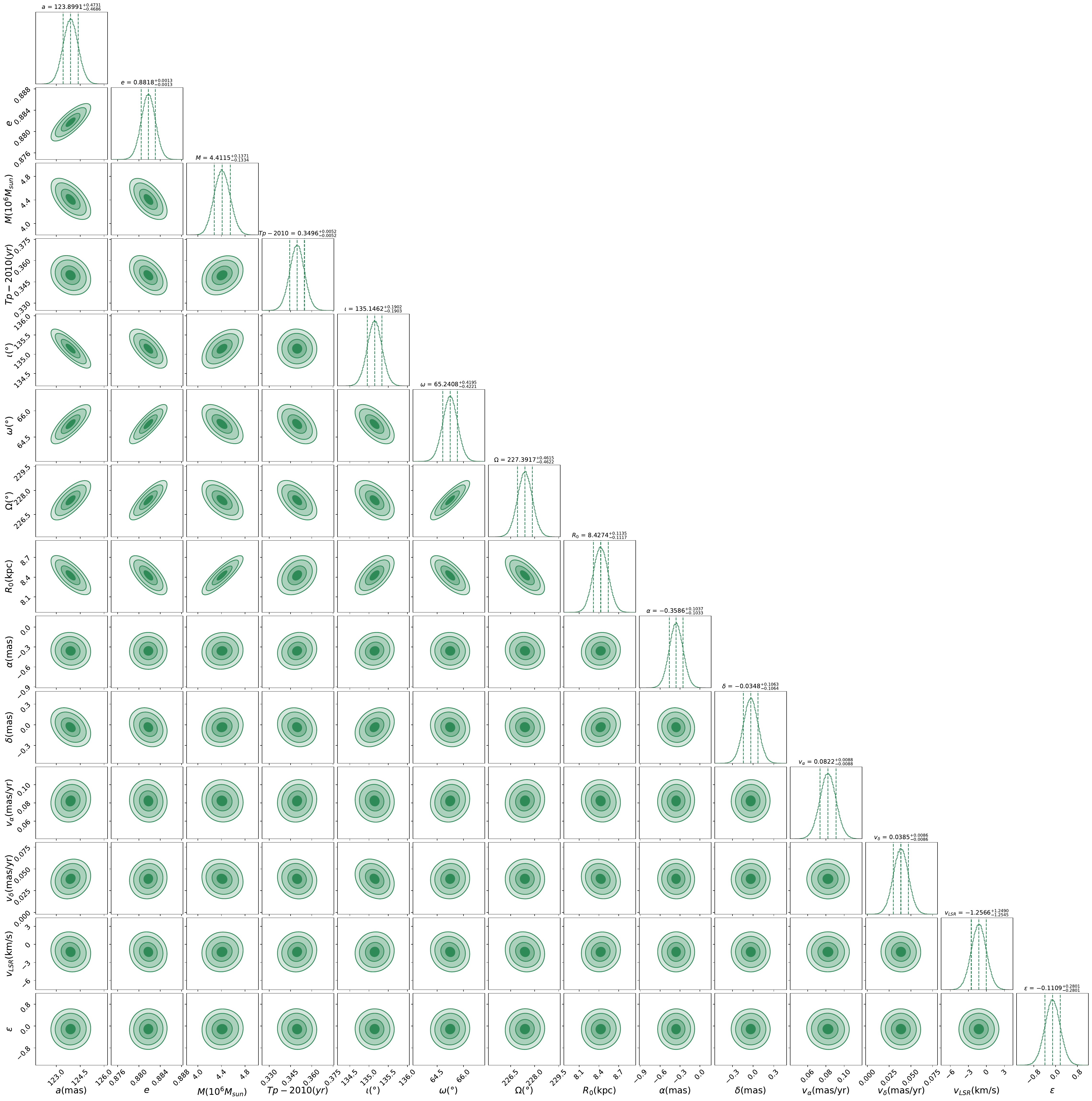}
\caption{
The posterior distribution of the orbital parameters of the S2 star and the deviation parameter $\epsilon$ is derived from the asymptotically flat Buchdahl-inspired metric. The three vertical bars represent confidence intervals of 16\%, 50\%, and 84\% from left to right. This corner plot provides insights into the correlation between each parameter through the contour map. A flatter contour map indicates a stronger correlation. Notably, the deviation parameter $\epsilon$ exhibits weak correlations with all other parameters. }\label{corner}
\end{figure*}

With all preparations made, we performed the MCMC analysis with the orbital data of S2 star at the Galatic center to constrain the deviation parameter $\epsilon$ in the Buchdahl-inspired spacetime, and the priors are given in Table.~\ref{priors}. The results of the posterior distributions of both the orbital parameters and the deviation parameter of Buchdahl-inspired spacetime are illustrate in Fig.~\ref{corner}. In this figure the vertical dashed lines are added to delineate different confidence intervals at 16\%, 50\%  and 84\% from left to right. The posterior distribution of all parameters converges well, and the orbital parameters of the S2 star align with general expectations. 

\begin{table}
\caption{\label{priors}%
Priors used in our MCMC analysis. We use uniform priors for the deviation function $\epsilon$ and the orbital parameters \{ $ M, R_0, a, e, \iota, \omega, \Omega, t_p   $\}. Those parameters \{ $\alpha, \delta, v_{\alpha}, v_{\delta}, v_{LSR}$ \}  used to describe the zero-point offset and the drift of the reference frame, are independent with gravity theory. Therefore we use the Gaussian priors  centered on the best values given in \cite{GRAVITY2}. Units: $M_{\odot}$ for solar mass, kpc for kiloparsec, mas for milliarcsecond, $^{\circ}$ for degree, and yr for year. }
\begin{ruledtabular}
\begin{tabular}{lcccl}
&\multicolumn{2}{c}{Gaussian priors} &\multicolumn{1}{c}{Uniform prior} & Estimate \\
\cline{2-3}  \cline{4-4}
 Paramaters& $\mu$ & $\sigma$ & - &results(95\% CL) \\
  \colrule
     $ M \;(10^{6}M_{\rm \odot})$ &   - & -  & [3, 5]  & $ 4.4115^{+0.2705}_{-0.2563}$  \\
     $R_{0}$ (kpc) & -  & - & [7, 10] & $ 8.4274^{+0.2240}_{-0.2182}$  \\
     $a$ (mas) & -    & - & [110, 150] & $123.899^{+0.9467}_{-0.9228}$ \\
     $e$ & -   & - & [0.8, 0.95] & $ 0.8818^{+0.0026}_{-0.0026}$  \\
     $\iota$ ($^{\circ}$)  & -    & -  & [120, 150] & $ 135.1462^{+0.3792}_{-0.3791}$ \\
     $\omega$ ($^{\circ}$) & - & - &  [50, 80] & $ 65.2408^{+0.8332}_{-0.8399}$ \\
     $\Omega$ ($^{\circ}$)  & - & - & [210, 240] & $ 227.3971^{+0.9179}_{-0.9174}$ \\
     $t_{\rm P}$ (yr) &  -  & - & [2008, 2012]  & $ 2010.3496^{+0.0103}_{-0.0103}$\\
     $\alpha$ (mas) & -0.9   & 0.15  & - & $ -0.3586^{+0.2051}_{-0.2049}$ \\
     $\delta$ (mas) & 0.07   & 0.12  & - & $ -0.0348^{+0.2097}_{-0.2112}$\\
     $v_{\rm \alpha}$ (mas/yr) & 0.08   & 0.01 & -  & $ 0.0822^{+0.0176}_{-0.0175}$\\
     $v_{\rm \delta}$ (mas/yr) & 0.0341  & 0.0096 & - & $ 0.0385^{+0.0170}_{-0.0172}$ \\
     $v_{\rm  LSR}$ (km/s) & -1.6 & 1.4 & - & $ -1.2566^{+2.4896}_{-2.4819}$\\
     \colrule
$\epsilon$ & - & - &[-1, 1] & $-0.1109^{+0.5561}_{-0.5581}$ \\
\end{tabular}
\end{ruledtabular}
\end{table}

\begin{figure}[h]
\centering
\includegraphics[width=8.1cm]{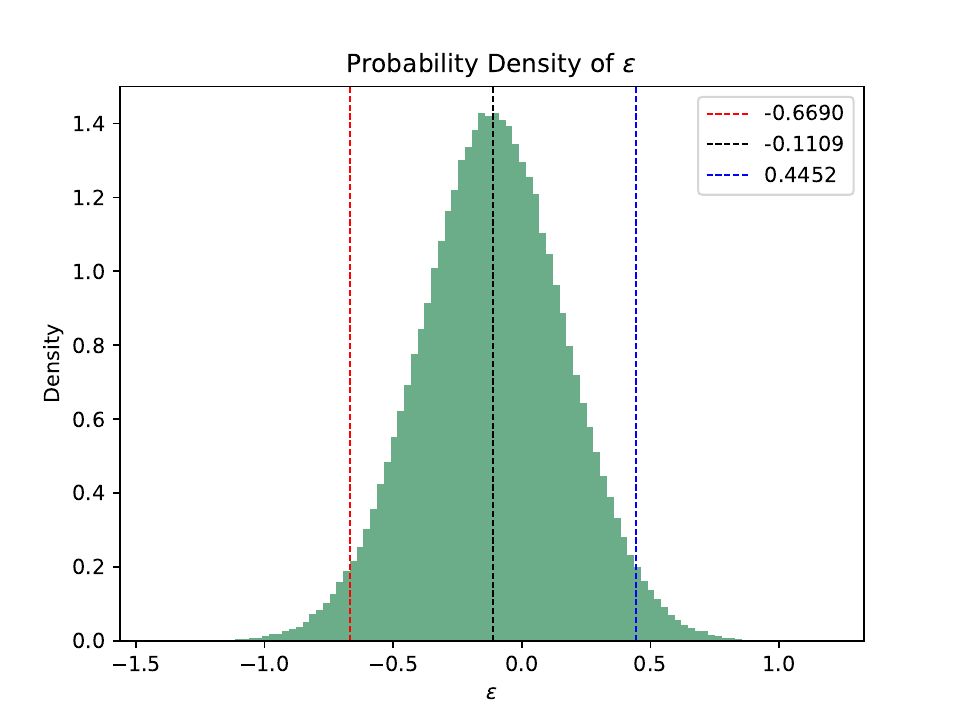}
\caption{The probability density distribution of the deviation parameter $\epsilon$, derived from the Buchdahl-inspired metric, is depicted in the plot. The vertical lines in red, black, and blue correspond to the lower limit, the peak, and the upper limit of $\epsilon$ under 95 \% confidence level, respectively. These lines mark key percentiles in the distribution, providing insights into the range and central tendency of the parameter.}\label{epsilon}
\end{figure}

In order to estimate the observational constraint on the deviation parameter $\epsilon$ in the Buchdahl-inspired spacetime, we plot the marginalized posterior distribution of $\epsilon$ in  Fig.~\ref{epsilon}. The vertical lines in red, black, and blue correspond to the lower limit, the peak, and the upper limit of $\epsilon$ under 95\% confidence level, respectively. Then the bounds on $\epsilon$ can be calculated from the corresponding posterior distribution of $\epsilon$. We found that the deviation parameter $\epsilon$ is distributed around zero, with its peak very close to zero (Fig. \ref{epsilon}), which is consistent to the prediction of the Schwarzschild black hole. Then we can place a bound on $\epsilon$ at 95\% confidence level as
\bqn
-0.6690 < \epsilon < 0.4452,
\eqn
with $\epsilon_{\rm peak} = -0.1109$. Comparing this result with the prediction obtained in (\ref{prediction}), the range of our MCMC result encompasses the predicted range, and they are on the same order of magnitude. This result is anticipated and well-supported, given that the GRAVITY Collaboration leveraged a more extensive set of observational data, encompassing various data types and sources. The comprehensive nature of their dataset contributes to the robustness and reliability of the obtained results.

\section{CONCLUSION}
In this paper, we employed the asymptotically flat Buchdahl-inspired metric as the spherically symmetric vacuum solution of ${\cal R}^{2}$ gravity. We identify a deviation parameter $\epsilon$ as a modification to the well-known Schwarzschild solution. By analyzing the motion of massive objects in this spacetime through perturbative dynamics, we observe that only the argument of pericenter $\omega$ undergoes a secular change given by $\Delta \omega = \frac{6\pi M}{a(1-e^2)}\Big(1-\frac{2}{3} \epsilon \Big).$
Comparing this with the precession of S2's orbit as measured by the GRAVITY Collaboration, we estimate the value of $\epsilon$ within the range $-0.42 < \epsilon < 0.134$.

Next, we utilize publicly available data for the S2 star in the Galatic center, including astrometric positions and radial velocity, and the precession. Through a Bayesian analysis with the \textit{emcee} Python package, we constrain the value of $\epsilon$. Our results indicate an expected range of $-0.6690 < \epsilon < 0.4452$, acknowledging that the GRAVITY Collaboration employed a more extensive dataset.

It is essential to note that this constraint, while insightful, currently falls short compared to tests within our solar system. Nevertheless, the S2 star's orbit around the galactic center would provide a meaningful test for the regime of strong gravitational fields, which might offer clues towards modifications beyond GR. We anticipate that with advancements in scientific and technological capabilities, the precision of Milky Way observations will continue to improve and prolonged. In the near future, these observations could offer more stronger bounds and enhance the accuracy of our analysis.

\section*{Acknowledgments}

This work is supported by the Zhejiang Provincial Natural Science Foundation of China under Grants No. LR21A050001 and No. LY20A050002, the National Natural Science Foundation of China under Grants No. 12275238, the National Key Research and Development Program of China under Grant No. 2020YFC2201503, and the Fundamental Research Funds for the Provincial Universities of Zhejiang in China under Grant No. RF-A2019015.

\end{document}